\documentclass[12pt,preprint]{aastex}
\usepackage{psfig,epsfig,amsfonts,amsmath,amssymb,graphicx,morefloats,appendix,tensor}
\usepackage{color}
\usepackage{natbib}
\begin{document}

\slugcomment{Accepted to ApJ: January 27, 2020}

\title{Properties of M Dwarf Flares at Millimeter Wavelengths 
}

\author{Meredith A. MacGregor\altaffilmark{1,2,3}, Rachel A. Osten\altaffilmark{4,5}, A. Meredith Hughes\altaffilmark{6}}

\altaffiltext{1}{Department of Terrestrial Magnetism, Carnegie Institution for Science, 5241 Broad Branch Road NW, Washington, DC 20015, USA}
\altaffiltext{2}{Department of Astrophysical and Planetary Sciences, University of Colorado, 2000 Colorado Avenue, Boulder, CO 80309, USA}
\altaffiltext{3}{NSF Astronomy and Astrophysics Postdoctoral Fellow}
\altaffiltext{4}{Space Telescope Science Institute, Baltimore, MD  21218 USA}
\altaffiltext{5}{Center for Astrophysical Sciences, Johns Hopkins University, Baltimore, MD 21218, USA}
\altaffiltext{6}{Department of Astronomy, Wesleyan University, Van Vleck Observatory, 96 Foss Hill Drive, Middletown, CT 06457, USA}

\begin{abstract}

We report on two millimeter flares detected by ALMA at 220~GHz from AU~Mic, a nearby M dwarf. The larger flare had a duration of only $\sim35$~sec, with peak $L_{R}=2\times10^{15}$~erg~s$^{-1}$~Hz$^{-1}$, and lower limit on linear polarization of $|Q/I|>0.12\pm0.04$.  We examine the characteristics common to these new AU Mic events and those from Proxima~Cen previously reported in MacGregor et al. (2018) -- namely short durations, negative spectral indices, and significant linear polarization -- to provide new diagnostics of conditions in outer stellar atmospheres and details of stellar flare particle acceleration. The event rates ($\sim20$ and $4$ events~day$^{-1}$ for AU~Mic and Proxima~Cen, respectively) suggest that millimeter flares occur commonly but have been undetected until now.  Analysis of the flare observing frequency and consideration of possible incoherent emission mechanisms confirms the presence of MeV electrons in the stellar atmosphere occurring as part of the flare process.  The spectral indices point to a hard distribution of electrons.  The short durations and lack of pronounced exponential decay in the light curve are consistent with formation in a simple magnetic loop, with radio emission predominating from directly precipitating electrons.  We consider the possibility of both synchrotron and gyrosynchrotron emission mechanisms, although synchrotron is favored given the linear polarization signal.  This would imply that the emission must be occurring in a low density environment of only modest magnetic field strength.  A deeper understanding of this newly discovered and apparently common stellar flare mechanism awaits more observations with better-studied flare components at other wavelengths. 

\end{abstract}


\section{Introduction}
\label{sec:intro}

Solar and stellar flares release energy across the entire electromagnetic spectrum \cite[e.g.,][]{Klein:2017}.  In the generally accepted picture of stellar flaring, magnetic field lines reconfigure in the outer stellar atmosphere and liberate energy.
Some of the energy from these reconnection events accelerates electrons that spiral along magnetic field lines, producing non-thermal emissions, predominantly incoherent and coherent radio emission and bremsstrahlung hard X-ray emission.  Electrons impact and heat ambient gas, resulting in prompt visible/UV flares from the chromosphere and longer-duration soft X-ray flares from the corona \citep{fisher1985}.  Since each wavelength probes different aspects of a flare, no single one gives a complete picture of the physics at work.  While non-thermal hard X-ray emission is commonly used to diagnose particle acceleration in solar flares, conclusive detections in stellar flares remain elusive \citep{Osten:2016}. Radio wavelength observations thus offer the only method to directly observe energetic particles in stellar flares \citep{bastian1998}. By probing accelerated electrons, these wavelengths enable us to test how the energy budget of accelerated particles versus thermal flare emission compares to the Sun \citep{Osten:2017}. 

Traditionally, most observational effort to characterize stellar radio emission has focused on microwave frequencies \citep{Dulk:1985,benzgudel2010}, due to the localization capability of interferometers coupled with available instrumental sensitivity.  The general interpretation of solar and stellar microwave emission has been gyrosynchrotron emission from a population
of mildly relativistic electrons, although bursts of highly circularly polarized radiation point to additional coherent emission mechanisms sporadically appearing \citep[e.g.,][]{Slee:2003}.  This frequency range suffers from a mixture of optically thick and thin gyrosynchrotron emission, however, which hampers interpretation and derivation of astrophysical parameters.   These electrons are accelerated high in the corona, and then experience injection onto magnetic field lines leading away from the acceleration site.  Magnetic loops trap particles due to the conservation of the magnetic moment \citep{benz2002}. Coronal magnetic loops also act like a dispersive element, with different parts of the electron distribution emitting at different locations within the magnetic loop \citep{bastian1998}.  Some of these particles may precipitate from the trap due to pitch angle scattering; both trapped and precipitating particles produce radio emission \citep{lee2002}. 

Variable stellar emission at millimeter wavelengths on timescales of seconds to minutes is also expected to occur due to the action of accelerated particles in a magnetoactive plasma.  This incoherent non-thermal emission indicates energetic MeV electrons \citep{white1992}.  Only a few stellar millimeter flares have been reported previously at frequencies around 100 GHz -- V773 Tau \citep{Massi:2006}, $\sigma$ Gem \citep{Brown:2006}, UX Ari \citep{Beasley:1998}, and GMR-A \citep{bower2003}.  Magnetic reconnection events producing flaring millimeter-wave emission have also been reported in the young accreting binary DQ Tau \citep{salter2010}, although \citet{tofflemire2017}'s study of optical flux maxima at periastron passage suggest that temporary accretion increases may also be occurring. More recently, \citet{mairs2019} reported on a very bright submillimeter flare in a binary T Tauri system which they interpreted as gyrosynchrotron/synchrotron radiation.  Recent observations with the Atacama Large Millimeter/submillimeter Array (ALMA) unexpectedly detected flaring emission from Proxima Cen at millimeter wavelengths \cite[233~GHz,][]{macgregor2018}.  These observations indicate optically thin emission with linear polarization during the flare, and open an entirely new window to study particle acceleration during stellar flares.   

Detailed modelling of white-light M dwarf flares reveals difficulties in the applicability of a solar-like prescription of accelerated particles in the standard solar flare model.  \citet{allred2006} performed radiative hydrodynamic simulations of how an M dwarf atmosphere responds to the input of a solar-like beam of electrons.  An earlier paper showed the appplicability of the approach to reproducing solar flare observations \citep{allred2005}, and some aspects of M dwarf flares were reproduced in the study, such as line emission and observed velocity shifts and Stark broadening.  The major failing of the solar model was the inability to replicate the magnitude of the observed continuum enhancements in M dwarf flares \citep{allred2006}.  The solar-like prescription of an electron beam cannot penetrate deeply enough into the stellar photosphere to provide direct heating that would result in the observed continuum emission, and backwarming from the X-ray -- EUV emission cannot provide the required energy input.  More recently, \citet{kowalski2015} have demonstrated through modelling of white-light M dwarf flares that a higher flare heating level in non-thermal electrons is required than typically found in solar flares to reproduce the observed color temperature and Balmer jump ratios self-consistently.  Additional studies are investigating how the distribution of accelerated particles affects white-light and radio measurements (A. Kowalski, priv. comm.).  These results are spurring effort to understand the implications for radio diagnostics of accelerated electrons.

Here, we report on the detection of millimeter flaring emission from AU Mic, another M dwarf star (M1V), with ALMA.  A member of the the $23\pm3$~Myr-old $\beta$ Pictoris moving group \citep{Mamajek:2014}, AU Mic has long been of interest to the exoplanet community.  It is one of only a handful of M dwarf stars known to host a debris disk, composed of the leftover material from the planet formation process, and has been imaged multiple times with ALMA \citep{MacGregor:2013,Daley:2019}.  AU Mic was already well-known as a flare star, and a previous flaring event was detected by the VLA at 9~mm \cite[34~GHz,][]{MacGregor:2016a}.  In addition, its coronal activity has been studied using high energy observatories \cite[notably][]{monsignorifossi1996,katsova1999}.  This new detection of another M dwarf flare with ALMA suggests that millimeter emission might be a common occurrence during stellar flares that has been missed by observations until now.  In \S\ref{sec:obs} and \S\ref{sec:aumic_flare} of this paper, we present analysis of archival ALMA data aimed at imaging the disk that also includes a significant flaring event from the star.  Then, in \S\ref{sec:discussion}, we discuss the properties of the full sample of millimeter flaring events detected from M dwarfs (AU Mic and Proxima Cen) and consider possible emission mechanisms.  \S\ref{sec:conclusions} presents our conclusions.

\section{New ALMA Observations of AU Mic}
\label{sec:obs}

AU Mic was observed with the ALMA 12-m array at 222~GHz (1.35~mm) three times on 2014 March 26, 2014 August 18, and 2015 June 24 (PI: Hughes, 2012.1.00198.S).  The original analysis and description of these observations are presented in \cite{Daley:2019}.  For all three scheduling blocks (SBs), there were $32-37$ antennas in the array spanning baseline lengths of $12-1320$~m.  The SB executed on 2015 June 24, which we consider in this paper, had 37 antennas with baselines between $30-1320$~m.  During this SB, AU Mic was observed in 6.5~min integrations or `scans' alternating with a phase calibrator, J2056-3208, for a total of $33$~min on-source.  For absolute flux calibration and bandpass calibration, Titan and J1924-2914 were used, respectively.  The weather was very good during this final observation with precipitable water vapor (PWV) ranging between $0.67-0.74$~mm. 

The correlator set-up for these observations was designed to achieve the original science goal of resolving the vertical structure of the edge-on AU Mic debris disk.  Three continuum spectral windows were defined at central frequencies of 213.5, 216.0, and 228.5~GHz with a total bandwidth of 2~GHz each.  One spectral window was centered on the CO J$=2-1$ line at 230.538001~GHz with a total bandwidth of 1.875~GHz.  The raw ALMA data were reduced in \texttt{CASA} \cite[][version 4.4]{Petry:2012} using the scripts provided with the data.  To invert the millimeter visibilities and produce deconvolved images, we used the \texttt{clean} task in \texttt{CASA} (version 4.7.0).

\section{Detection of a Millimeter Flare from AU Mic}
\label{sec:aumic_flare}

\begin{figure}[t]
\begin{minipage}[h]{0.4\textwidth}
  \begin{center}
       \includegraphics[scale=0.75]{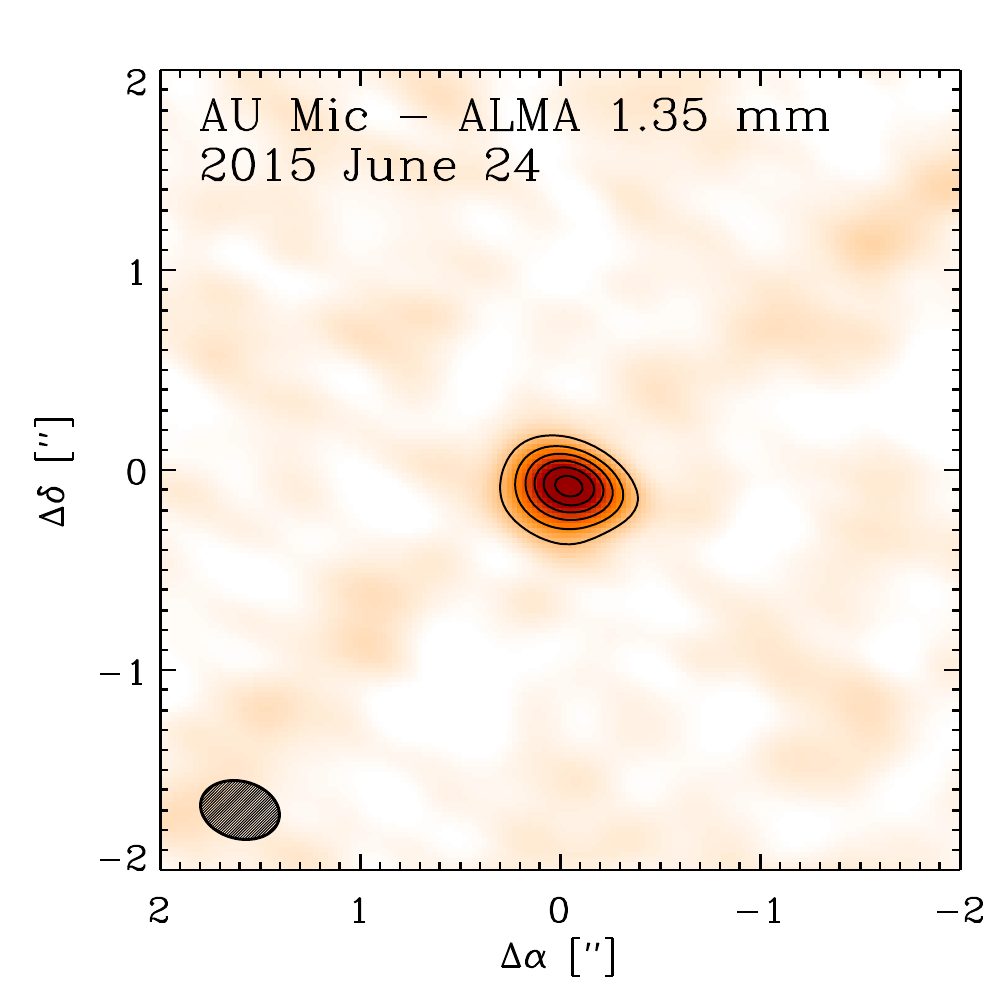}
  \end{center}
 \end{minipage}
\begin{minipage}[h]{0.6\textwidth}
  \begin{center}
       \includegraphics[scale=0.6]{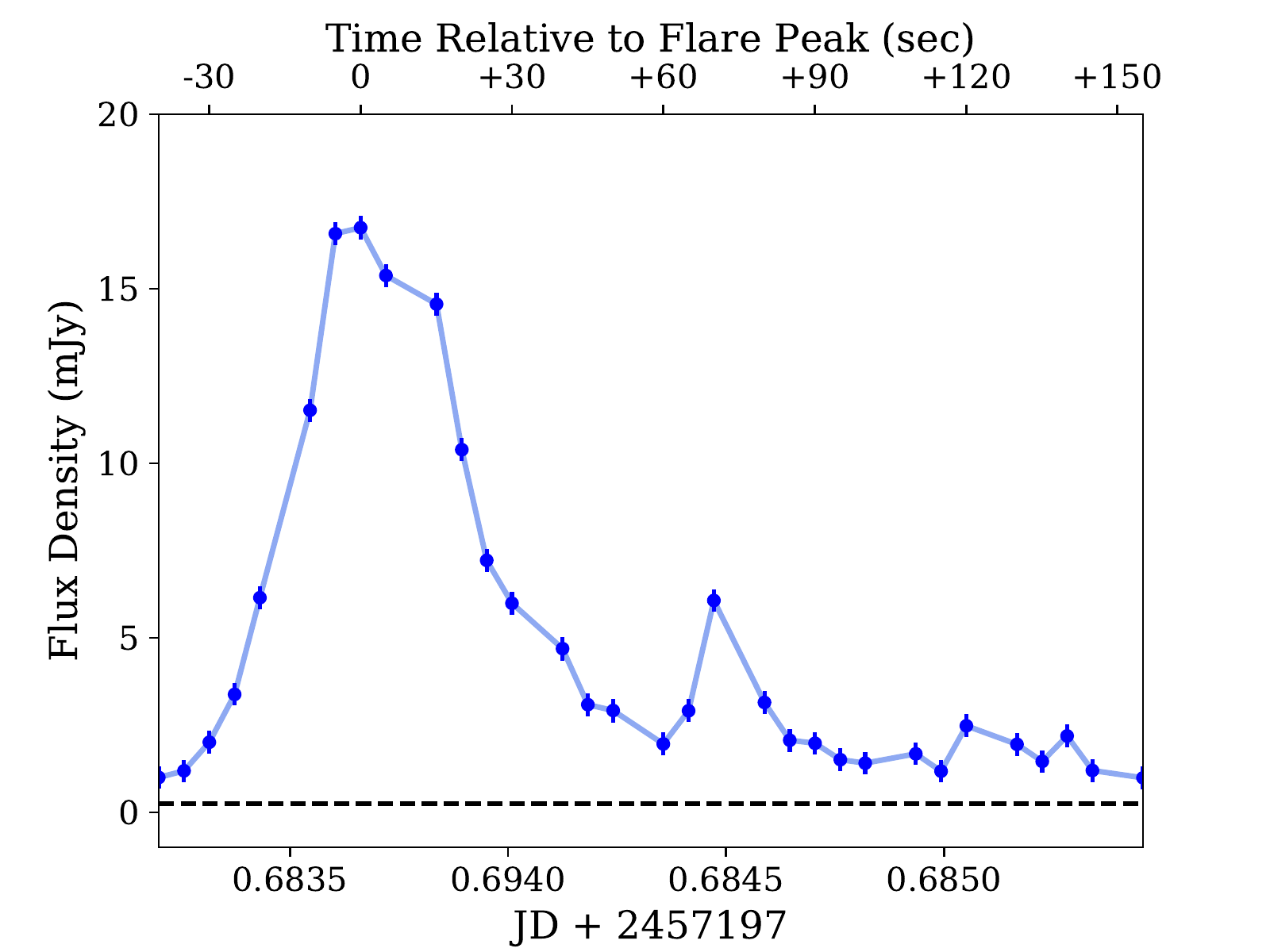}
  \end{center}
 \end{minipage}
\caption{\small We detect a large flaring event from AU Mic in the 2015 June 24 ALMA data.  \emph{(left)} An image of AU Mic during the $\sim3$~min flaring event with all data averaged together.  In order to isolate the emission from the star, we only include baselines $>100$~k$\lambda$.  The synthesized beam size is $0\farcs42 \times 0\farcs30$ with natural weighting (shown by the ellipse in the lower left corner).  Contours are in increments of $5\times$ the rms noise of $0.08$~mJy~beam$^{-1}$.  
\emph{(right)}  The light curve of the flaring event shows an initial large flare with a peak flux density of $16.8\pm0.4$~mJy followed by a smaller flare with a peak flux density of $6.07\pm0.35$~mJy.  All flux densities were determined by fitting point source models to the millimeter visibilities in 5~sec intervals using the \texttt{uvmodelfit} task in \texttt{CASA}. The dashed black line indicates the $3\sigma$ detection threshold of $0.24$~mJy for these observations. 
}
\label{fig:image}
\end{figure}

We detect a significant flaring event from AU Mic at the beginning of the 2015 June 24 ALMA observations.  Figure~\ref{fig:image} (left panel) shows a natural weight (synthesized beam size of $0\farcs42 \times 0\farcs30$) image of AU Mic during this event.  A central point source is detected at $32\sigma$.  In order to isolate the stellar emission from disk emission, we only image baselines $>100$~k$\lambda$.  Although this process does not completely exclude the edge-on debris disk, no emission is detected from the disk at $>3\sigma$ in the image.  No other flares are detected during the 2014 March 26 or August 28 observations.  However, a point source is still detected at the stellar position during quiescent periods with a flux that appears to vary on month- to year-long timescales.  \cite{Daley:2019} fit for the quiescent stellar flux density during each of the three observations and obtain values of 390, 150, and 220~$\mu$Jy for the 2014 March 26, 2014 August 18, and 2015 June 24 observations, respectively.  Early ALMA observations from 2012 (PI: Wilner, 2011.0.00142.S) also detected a central point source with a flux density of 320~$\mu$Jy, although no large flares were seen \citep{MacGregor:2013}.

In order to fully explore the time variable nature of the 2015 June 24 event, we fit point source models to the millimeter visibilities in 5~sec intervals using the \texttt{uvmodelfit} task in \texttt{CASA}.  As with imaging, we exclude baselines $<100$~k$\lambda$ to isolate the stellar emission.  The resulting light curve is shown in Figure~\ref{fig:image} (right panel).  At the start of the event, the star brightens suddenly over $\sim30$~sec to reach a peak flux density of $16.8\pm0.3$~mJy.  Assuming a distance to the source of $9.725\pm0.005$~pc from \emph{Gaia} \citep{Gaia:2018}, this yields a peak luminosity of 
$1.96\pm0.04\times10^{15}$~erg~s$^{-1}$~Hz$^{-1}$, a factor of $\sim10\times$ brighter than the recently detected Proxima Cen millimeter flare with a peak luminosity of $2.04\pm0.15\times10^{14}$~erg~s$^{−1}$~Hz$^{-1}$ \citep{macgregor2018}.  Following the peak, the brightness declines for $\sim60$~sec before briefly increasing again to a maximum of $6.1\pm0.4$~mJy and then falling off to quiescent values.  All together, the entire AU Mic flaring event lasts for $\sim3$~min.  The observed sharp rise and gradual decline in brightness are similar to what was seen for the Proxima Cen flare and is characteristic of stellar flares at other wavelengths.  We note that no variability is seen in the gain calibrator flux densities (interleaved with the source observations) on similar timescales.

\begin{figure}[t]
\begin{minipage}[h]{0.5\textwidth}
  \begin{center}
       \includegraphics[scale=0.55]{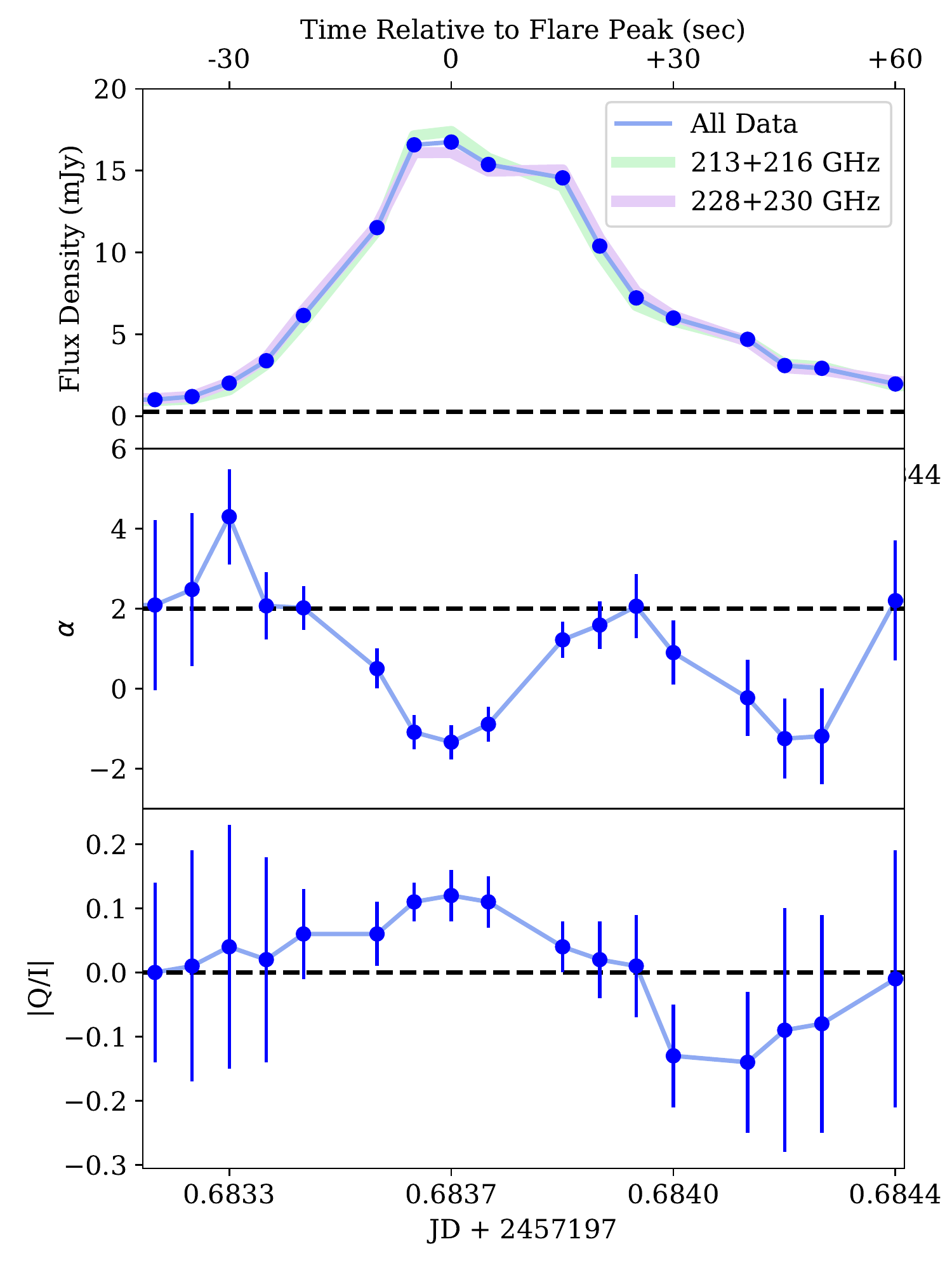}
  \end{center}
 \end{minipage}
\begin{minipage}[h]{0.5\textwidth}
  \begin{center}
       \includegraphics[scale=0.55]{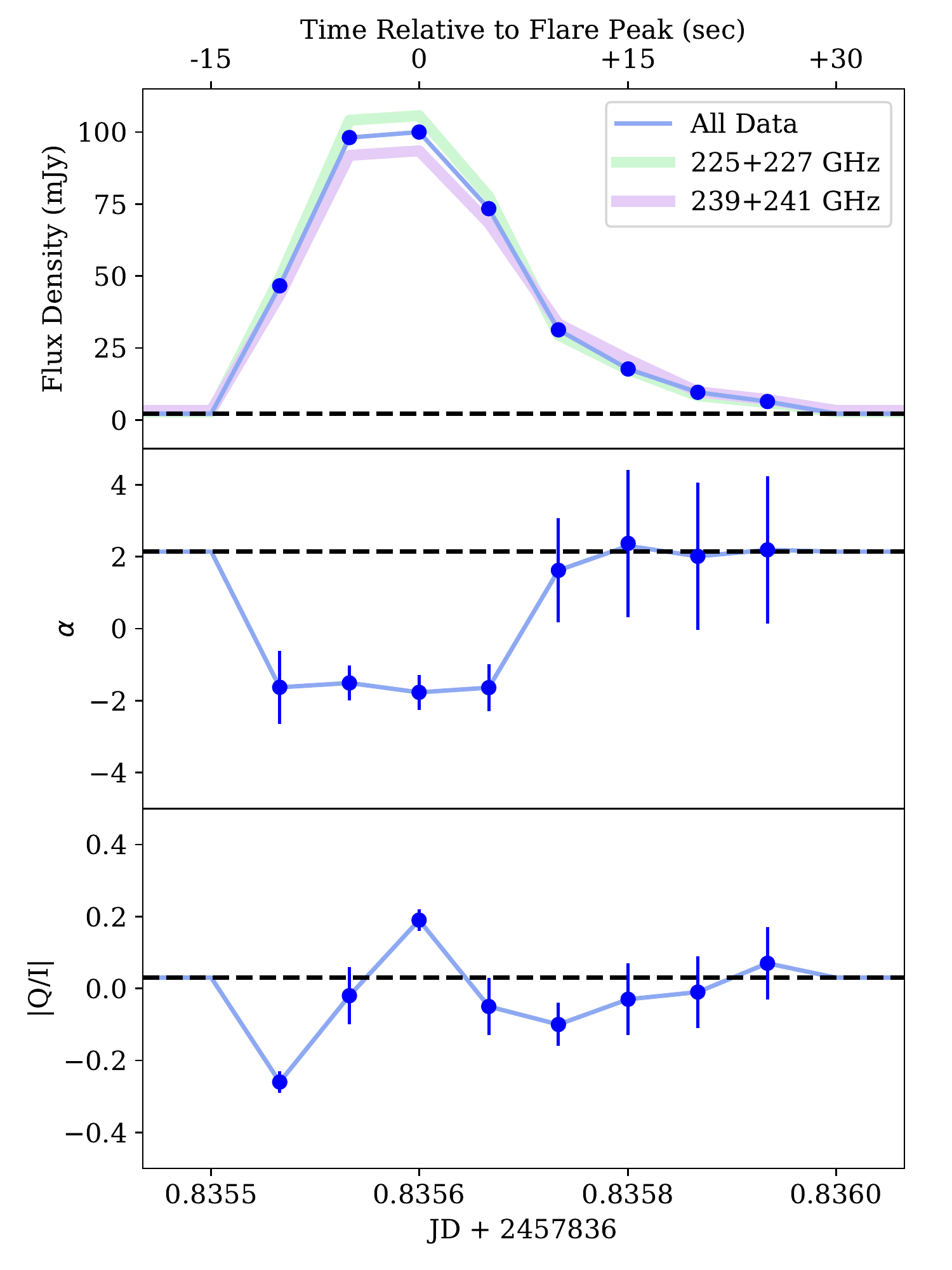}
  \end{center}
 \end{minipage}
\caption{\small The large flares from both AU Mic \emph{(left)} and Proxima Cen \emph{(right)} show similar properties.  Here, the flux density \emph{(top)}, spectral index, $\alpha$, where $F_\nu \propto \nu^\alpha$ \emph{(middle)}, and lower limit on the fractional linear polarization, $|Q/I|$, \emph{(bottom)} are shown for both events.  To determine these properties, we fit point source models to the millimeter visibilities in 5~sec intervals using \texttt{uvmodelfit} in \texttt{CASA}.  In all panels, the horizontal dashed line indicates the quiescent (non-flaring) value of that parameter.  In the top panel, we plot the combined flux density (blue) along with the flux density determined for the lower (green) and upper (purple) sidebands independently.  
}
\label{fig:prop}
\end{figure}

By default, ALMA continuum observations include two linear polarizations (XX and YY) and 8~GHz of frequency bandwidth split up into four spectral windows.  For the AU Mic observations discussed here, the spectral windows span a total frequency range of $17$~GHz between 213.5 and 230.5~GHz.  We can use this extra information to determine both the spectral index as a function of frequency ($\alpha$, where $F_\nu \propto \nu^\alpha$) and a lower limit on the fractional linear polarization ($|Q/I|$) during the flare.  Figure~\ref{fig:prop} shows the flux density, spectral index, and lower limit on the fractional linear polarization binned in 5~sec intervals for both the AU Mic (left) and Proxima Cen (right) flares.  We note that $|Q/I|$ is only a lower limit to the true linear polarization fraction $p_\text{QU}^2 = (Q/I)^2 + (U/I)^2$ since we do not constrain Stokes $U$ with these observations.  Both millimeter flares show similar characteristics -- at peak, the spectral index becomes steeply negative while the fractional linear polarization increases.  To calculate the spectral index for the AU Mic event, we fit point source models independently to the millimeter visibilities from the lower (213.5 and 216~GHz spectral windows) and upper (228.5 and 230.5~GHz spectral windows) sidebands.  In order to check the accuracy of the frequency-dependent amplitude calibration performed by the ALMA pipeline, we compared the results for the calibration sources to the ALMA calibrator catalogue \citep{Bonato:2018}.  At peak, the flux densities are $17.4\pm0.4$~mJy and $16.1\pm0.4$~mJy for the lower and upper sidebands, respectively, yielding a spectral index of $\alpha = -1.3\pm0.4$.  If we fit point source models to the XX and YY polarizations independently, we obtain flux densities of $F_\text{XX}=17.0\pm0.4$~mJy and $F_\text{YY}=15.1\pm0.4$~mJy at peak.  The magnitude of Stokes $Q = \langle E_\text{X}^2\rangle - \langle E_\text{Y}^2 \rangle$ as a fraction of Stokes $I = \langle E_\text{X}^2\rangle + \langle E_\text{Y}^2 \rangle$ yields $|Q/I| = 0.12\pm0.04$.  Here, $E_\text{X}$ and $E_\text{Y}$ are  antenna voltage patterns.  Outside of flaring, both AU Mic and Proxima Cen exhibit spectral indices $\sim2$ and no linear polarization, consistent with blackbody emission from a quiescent stellar photosphere in the Rayleigh-Jeans regime.

\begin{figure}[t]
  \begin{center}
       \includegraphics[scale=0.58]{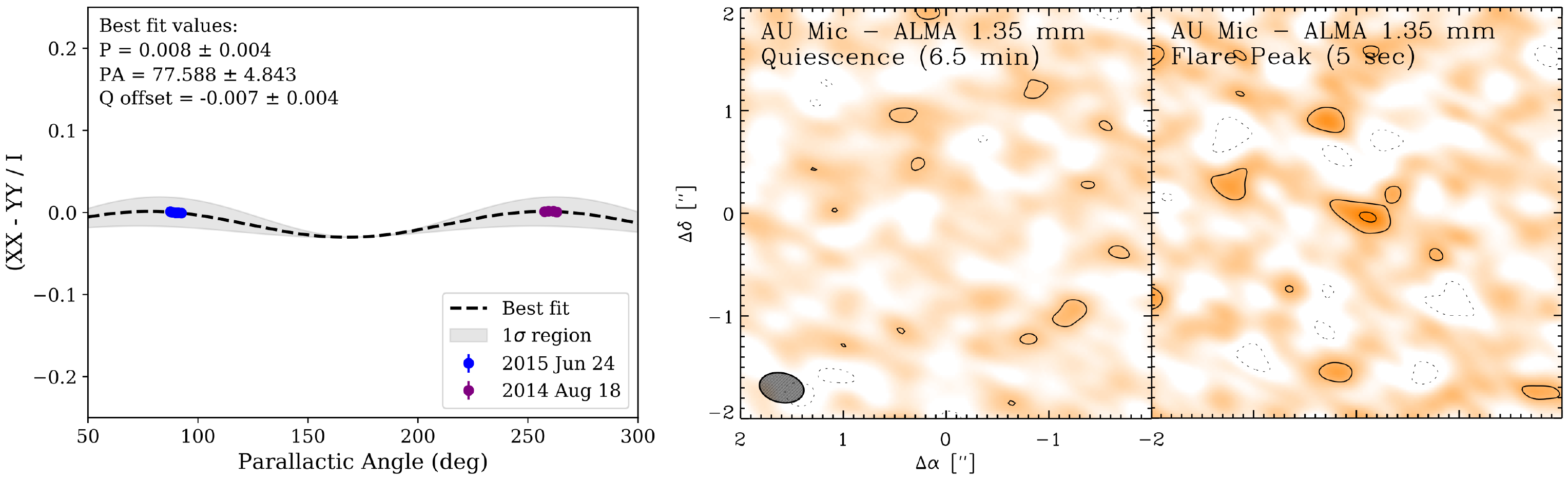}
  \end{center}
\caption{\small There is no indication of significant polarization from AU Mic during quiescence.  \emph{(left)} The $(XX-YY)/I$ intensity ratio as a function of parallactic angle.  Data from 2015 June 24 and 2014 August 18 are plotted in blue and purple, respectively (error bars are smaller than the symbol size).  The best-fit values and $1\sigma$ uncertainties for the polarization percentage (P), polarization position angle (PA), and constant normalized Q offset (Q offset) are given in the upper left corner and indicated on the plot by the dashed balck line and gray shaded region.  \emph{(center, right)} Natural weight XX-YY difference images of AU Mic during a 6.5-min quiescent scan and during a 5-sec integration at the flare peak.  As in Figure~\ref{fig:image}, we only include baselines $>100$~k$\lambda$ and the synthesized beam size is $0\farcs42 \times 0\farcs30$ (shown by the ellipse in the lower left corner).  Contours are in increments of $2\times$ the rms noise of $0.17$ and $0.54$~mJy~beam$^{-1}$ for the quiescent and flaring images, respectively.
}
\label{fig:pol}
\end{figure}

No polarization calibration was performed during either the AU Mic or Proxima Cen ALMA observations, and no cross products (XY and YX) are available.  However, it has been shown previously that polarimetric information can be recovered for ALMA dual polarization observataions \citep{Vidal:2016,Liu:2016}.  Since ALMA antennas have alt-azimuth mounts, a target source rotates with respect to the receiver frame over the course of an observation.  The fractional linear polarization, $|Q/I|$, can be defined as a function of the polarization percentage and parallactic angle of the target source:

\begin{equation}
    \left|\frac{Q}{I}\right| = \frac{XX-YY}{2I} = P \times \text{cos}(2(\psi - \eta)) + \delta \;\;\;,
\end{equation}

\noindent where $Q$ and $I$ are the observed Stokes Q and I fluxes, $\delta$ is a constant normalized $Q$ offset due to amplitude calibration errors or polarization leakage, $P$ is the polarization percentage, $\psi$ is the polarization position angle in the sky frame, and $\eta$ is the parallactic angle. Using this method, \cite{Vidal:2016} obtain similar results to full ALMA polarization observations of  3C286.  Since the AU Mic flare duration is too short for the parallactic angle to vary significantly, we cannot carry out this analysis for the flaring emission.  However, a larger spread in parallactic angle is available in quiescence.  We consider the observations taken on 2015 June 24 and 2014 August 18 that use the same array configuration, and determine $|Q/I|$ for each 6.5-min scan (roughly $2\deg$ in parallactic angle).  The results are shown in Figure~\ref{fig:pol} along with the best-fit values and $1\sigma$ uncertainties for the polarization percentage, position angle, and constant normalized Q offset obtained with a non-linear least squares fit.  The range of parallactic angle sampled by the observations and thus the constraints on these parameters are limited, but we can nonetheless conclude that almost no polarization is detected during quiescence.  The best-fit polarization percentage is $0.008\pm0.004$ with a $Q$ offset of $-0.007\pm0.004$, yielding a constraint on the effect of amplitude calibration errors and polarization leakage.  At face value, these results suggest that the instrumental contribution to any measured polarization is at the $\sim1\%$ level for these observations.  However, we note that ALMA instrumental polarization has been measured previously at the few percent level in some Band 6 observations. As a result, some caution must always be exercised when interpreting an observed polarization signal without appropriate calibration.  Figure~\ref{fig:pol} also shows $XX-YY$ difference images (with natural weighting) for AU Mic during a 6.5-min quiescent scan (center) and during a 5-sec interval at the flare peak (right).  No residuals are present during quiescence, but a $4\sigma$ peak is evident at the source position during the flare.  From this detailed analysis, we conclude that our detection of linear polarization during the AU Mic flaring event is robust despite the lack of dedicated polarization calibration.

\section{Discussion}
\label{sec:discussion}

Our new detection with ALMA of non-thermal, variable millimeter flaring at 222~GHz from AU Mic, a nearby M dwarf star, highlights the need to understand this phenomena and explore the impacts for both the star and any planets surrounding it.  In this section, we discuss the salient characteristics of AU Mic's flaring behavior at millimeter wavelengths and the implications for potential emission mechanisms.  We include in this discussion the millimeter flaring behavior of another M dwarf, Proxima Cen, as recently described by \citet{macgregor2018}, due to commonalities of behavior.  The flares we have detected at millimeter wavelengths from these two M dwarf stars have three unique characteristics: (1) short durations, (2) falling spectral energy distributions with frequency, and (3) evidence for linear polarization.  For the following analysis, we consider the complete sample of six millimeter flares detected by ALMA from M dwarf stars -- one large flare and one small flare detected from AU Mic, along with the one large flare and three small flares detected from Proxima Cen.  For the most general treatment, we first consider the observable properties of the millimeter
flares (\S\ref{sec:properties}), before proceeding to physical inferences about the flares and their environment derived in a manner that is agnostic about the emission mechanism (\S\ref{sec:flarechar}). Finally in \S\ref{sec:interpretation} we consider these effects together and discuss inferences about the emission mechanism.

\subsection{Properties of M Dwarf Millimeter Flares}
\label{sec:properties}

\subsubsection{Luminosity}
\label{sec:luminosity}

Table~\ref{tab:prop} lists the peak radio luminosity ($L_R$) for the six flaring events detected with ALMA to date, all of which are more luminous than millimeter flares detected previously from other sources.  The large burst on Proxima Cen had a peak radio luminosity of almost 2$\times$10$^{14}$~erg~s$^{-1}$~Hz$^{-1}$ \citep{macgregor2018}, about a factor of $10\times$ larger than the largest submillimeter flares seen on the Sun \citep{Krucker:2013}.  In our sample of six flare events, the burst on Proxima Cen is exceeded in peak radio luminosity by both events originating from AU~Mic.  Indeed, the largest AU Mic event has a peak luminosity another $10\times$ higher that the largest Proxima Cen flare.   The 90~GHz event on GMR-A (albeit a significantly different K5V star) seen by \citet{bower2003} had a peak radio luminosity of $\sim10^{19}$~erg~s$^{-1}$~Hz$^{-1}$.

\subsubsection{Event Occurrence}
\label{sec:occurrence}

We can estimate the occurrence rate of variable millimeter emission by examining the fraction of time each source was observed in an enhanced flux state relative to the total amount of observing time on-source.  For AU~Mic, the data reveal a span of about 2~min of flaring activity.  The on-source time in each of the three ALMA scheduling blocks for AU~Mic described here and in \citet{Daley:2019} is $33-35$~min, to which we add the 45~min of on-source time from an earlier ALMA observation of AU~Mic described by
\citet{MacGregor:2013} for a total monitoring time of 2.4~hours.  The percentage of time AU~Mic was observed in a flaring state is then about 1.4\%. Proxima Cen was observed by ALMA in scheduling blocks that spanned 23.4~hours, with 6.58~min on target in a loop that included target and phase calibrators \citep{macgregor2018}. Overall flux and bandpass calibration were also performed during this time, but we do not have a separate estimate of the total on-source time. The total span of flaring for Proxima Cen was also approximately 2~min, leading to an upper limit on the percentage of time flaring to be $<$0.1\%.
We use the expressions of \citet{Gehrels1986} to estimate 95\% confidence intervals for the small numbers of events detected and associated error on the event rate.  For AU Mic, the number of flares $n_{\rm fl}=2$ observed over a total of 2.4~hours, yielding an event rate of $\sim20$ events~day$^{-1}$ with a 95\% confidence level range of $3.6-63$ events~day$^{-1}$.  For Proxima Cen, $n_{\rm fl}=4$ over 23.4~hours, giving an event rate of 4 events~day$^{-1}$ with range $1.4-9.4$ events~day$^{-1}$.

\subsubsection{Temporal Behavior}
\label{sec:temporal}

\begin{figure}[t]
  \begin{center}
       \includegraphics[scale=0.7]{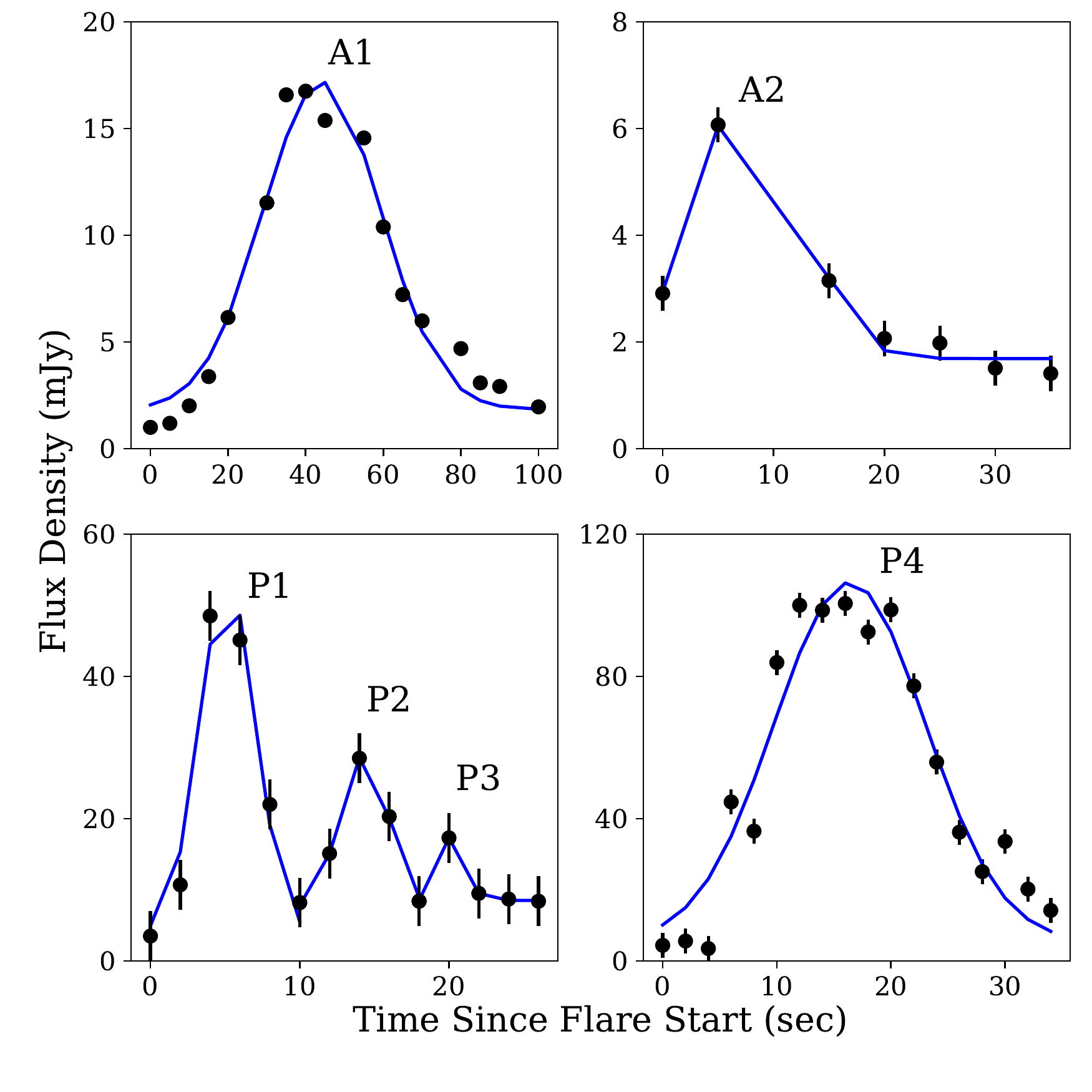}
  \end{center}
\caption{\small Light curves for all millimeter flares observed on AU~Mic \emph{(top)} and on Proxima Cen \emph{(bottom)}.  Each flare is labeled (A$n$ and P$n$ for AU Mic and Proxima Cen, respectively) and the key properties of all flares are listed in Table~\ref{tab:prop}.  The blue line in each panel shows a simple Gaussian fit to the data.
}
\label{fig:flarelc}
\end{figure}

Figure~\ref{fig:flarelc} shows the detailed temporal behavior of all six ALMA-detected flares. The duration of these flares are notably very short, shorter than typical flaring timescales observed at other wavelengths. Table~\ref{tab:prop} lists the peak flux densities and $t_{1/2}$ values for all six events.  Together, these events span a minimum $t_{1/2}$ value of just over 2~sec to a maximum $t_{1/2}$ value of 35~sec.  The temporal structure of these millimeter flares is mostly symmetric, with no pronounced exponential tail following the flare peak.  This is illustrated by the simple Gaussian fits over-plotted on the light curves in Figure~\ref{fig:flarelc}.  We note that the integration times for the ALMA data are only $\sim$5 seconds, and so there may be shorter fainter bursts occurring that we are not able to detect.

\subsubsection{Frequency Behavior}
\label{sec:frequency}

The flares discussed here were observed at central frequencies of 222 and 233~GHz for AU Mic and Proxima Cen, respectively, significantly higher than previously reported millimeter stellar flares.  For both of these M dwarfs, one large flare was observed along with at least one additional smaller enhancement either preceding or following within a few minutes.  Due to sensitivity limitations, we could only determine the spectral index for the two largest flares. Both the largest AU Mic (A1) and Proxima Cen (P4) flares exhibit negative spectral indices, with $\alpha_\text{A1}=-1.30\pm0.05$ and $\alpha_\text{P4}=-1.8\pm0.4$.  These values are indicative of a falling spectral energy distribution, which suggests optically thin emission for both events.  We note that in the absence of microwave or additional millimeter-wave coverage at lower frequencies, we cannot make any statements about the broad-band spectral energy distribution from microwave to millimeter wavelengths.

\subsubsection{Polarization Behavior}
\label{sec:polarization}

Surprisingly, both of the largest AU Mic and Proxima Cen millimeter flaring events (A1 and P4) show evidence for linear polarization during the rise phase.  There are no constraints on the presence of circular polarization during either of these M dwarf flares, so we cannot rule out a component of circular polarization.  There has only been one other report of linear polarization during a centimeter wavelength flare on the pre-main sequence star V773 Tau \citep{phillips1996}.  Along with subsequent studies, \cite{phillips1996} have investigated centimeter and millimeter-wavelength (90~GHz) variability from the V773 Tau system under the assumption of synchrotron emission.  Solar coronal studies have never revealed linear polarization at millimeter wavelengths.  The solar corona exhibits a large degree of Faraday rotation, and it is thought that differential Faraday rotation across the coronal source reduces the linear polarization as the emission propagates \citep{bastian1998}.

\subsection{Flare Characteristics Derived from the Observations}
\label{sec:flarechar}

\subsubsection{Flare Bolometric Energy Derived from Event Rate}
\label{sec:bolometric}

The occurrence rates derived in Section~\ref{sec:occurrence} can be related to a flare energy under the assumption that the particle acceleration event that results in these increased emissions is part of the flare process, and thus a millimeter flare should be accompanied by white-light flaring.  Then, by using an appropriate flare frequency distribution (FFD), we can estimate  a minimum bolometric flare energy consistent with this rate.  For AU~Mic, no FFDs specific to this star have been published, although separate observational campaigns in 2018 will remedy this (A. F. Kowalski, priv. comm., \& T. Barclay, priv. comm.).  FFDs for magnetically active, early M stars are also not prevalent, largely due to the relative paucity of such stars in the solar neighborhood \citep{west2004}.  We estimate a corresponding flare frequency using the FFD of inactive early M dwarfs from \cite[][Figure~8]{hawley2014}.  An event rate of $\sim20$ events~day$^{-1}$ yields an approximate minimum $E_{U}$ energy range of $10^{29}-10^{30}$~erg, corresponding to a bolometric value in the range of $10^{30}-10^{31}$~erg.  \citet{howard2018} published a cumulative FFD for Proxima Cen by combining two optical datasets (their Figure~3) and estimate a bolometric flare energy using the method of \citet{ostenwolk2015}.  Our estimated event rate of roughly 4 events~day$^{-1}$ corresponds to a minimum bolometric flare energy of a few times $10^{30}-10^{31}$~erg.

\subsubsection{Trapping vs. Precipitation Derived from Timescales and Light Curve Behavior}
\label{sec:trapping}

We assume that particles accelerated during these M dwarf millimeter flares follow magnetic field lines outlining loop-like structures.  Particles with pitch angles larger than a critical value become trapped in the magnetic loop \citep{benz2002}, while particles with lower pitch angle values precipitate directly.  Trapped particles eventually scatter into the loss-cone and precipitate from the trap, losing a significant fraction of their energy in the trap before they precipitate.  Both trapped and directly precipitating particles in a magnetic loop produce radio emission.  Trapped particles decay with a characteristic timescale related to the trapping process, $\tau$, which should be observable in the radio flare light curve. The time profile of the directly precipitating particles should reflect the time profile for injection into the trap, modulo any propagation effects in the atmosphere.  \citet{lee2002} utilized a model for the injection function and predicted microwave flux density variations as a function of frequency in a few solar flares, under varying assumptions of trapping and precipitation effectiveness (and type).  Their Figure~8 compares time varying injection for systems with good trapping (electron escape rate, $\nu \approx 10^{-3}$~s$^{-1}$) and good precipitation ($\nu \approx 10^{-1}$~s$^{-1}$).  In the case of good precipitation, the optically thin curves at the highest frequencies show a much faster decay with little evidence for an exponential tail.  Comparing the results of \citet{lee2002} with the symmetric nature of the light curves noted in Section~\ref{sec:temporal} suggests that trapping is not particularly effective in these millimeter flaring events.  

We do not have any independent constraints on the temporal profile of the injection function, but it must be less that the duration of these flares, which are at the short end of observed flare durations.  \citet{Kowalski:2013} reported durations for a handful of M dwarf flares in the blue-optical; while the median $t_{1/2}$ value is $\sim2.5$~min, less than one quarter of the sample had durations less than a minute.  More recently, \citet{Kowalski:2016} obtained median $t_{1/2}$ values of only 35.5~sec for blue-optical flares from M dwarfs. \citet{huenemoerder2010} noted much longer durations, spanning less than one ksec (or $\sim15$~min) to many tens of ksec, for a sample of coronal X-ray flares from the nearby flare star EV~Lac.   Only coherent emissions on M dwarf flare stars at centimeter wavelengths exhibit shorter duration bursting behavior \citep[as short as a few milliseconds, e.g.,][]{osten2008}.

\subsubsection{Densities \& Magnetic Field Strengths Derived from Linear Polarization Signals}
\label{sec:magnetic}

\citet{phillips1996} speculated on possible intrinsic and extrinsic explanations for the origin of linear polarization during the rise phase of a flare from V773 Tau, the only other stellar flare for which linear polarization has been detected.  The intrinsic possibilities considered were (1) a sudden injection of a new, more energetic electron population corresponding to the rise phase of the flare, and (2) the emergence of a polarized region from occultation by another optically thick component. The extrinsic possibilities were (3) highly efficient polarization of unpolarized radiation by a free electron Thomson backscattering halo, and (4) conversion of circular polarization to linear via a breakdown of the commonly assumed quasi-circular approximation for the electromagnetic wave.  In order to explain the observed linear polarization signal, the radio-emitting regions on V773 Tau needed to lie above the X-ray emitting regions,  making (1) the most likely explanation to encompass both the intensity rise as well as the linear polarization signal.  The intervening years of research have come to similar conclusions about non-cospatiality of X-ray and radio emission when sufficient information on size scale and geometry have been obtained \citep{favata2000,osten2006}.  This conclusion is somewhat inevitable when considering the propensity for X-ray emission to be maximized in compact dense structures, due to the $n_{e}^{2}$ influence of the radiation, and the energetic electrons producing radio emission surviving in low-density, extended magnetic structures.

The amount of linear polarization observed can place limits on the electron density and magnetic field  strengths involved in the flaring structures.  We use studies of internal Faraday rotation effects in transparent synchrotron sources \citep[e.g.,][]{cioffijones1980} to explore the parameter space of magnetic field strength and electron density in the source that could plausibly result in detection of a linearly polarized signal. We then
confront those with our observations of significant linear polarization 
in both the P4 and A1 flares.
%estimate the amount of Faraday rotation originating from a single magnetic loop using two different scenarios to explore the physical conditions that might be present, and confront those with our observations of significant linear polarization in both the P4 and A1 flares.  
To start, we take density and loop sizes derived from previously observed flares in the the EUV for AU Mic \citep{monsignorifossi1996,katsova1999} and the X-ray for Proxima Cen \citep{Gudel:2002}.  For the AU Mic flare, the electron density range is 3$\times$10$^{12}-$2$\times$10$^{13}$ cm$^{-3}$, with a size of 1-2 R$_{\star}$ \citep{katsova1999}.  \citet{Gudel:2002} determined electron density limits for Proxima Cen during the flare of 2$\times$10$^{10}-$4$\times$10$^{11}$ cm$^{-3}$ and loop half length about 10$^{10}$ cm.  The rotation measure is given by

\begin{equation}
    RM = 0.812 \;\; \text{rad}\; \text{m}^{-2} \times \int n_{e} (\text{cm}^{-3}) B_{||} (\mu\text{G}) dl (\text{pc}) \;\;\;,
\end{equation}

\noindent where $n_{e}$ is the electron density in cm$^{-3}$, $B_{||}$ is the parallel magnetic field in $\mu$G, and $dl$ the line of sight path length in pc.  Following the approach of \citet{phillips1996}, we take the line of sight path length to be one tenth the loop size and $B_{||}$ to be 10~G.  We use stellar radii of 0.83~R$_{\odot}$ for AU Mic \citep{white2015} and 0.14~R$_{\odot}$ for Proxima Cen \citep{angladaescude2016}.  
 For the observed wavelength of flaring emission (1.3~mm), even the lower limits on the electron density would result in Faraday rotation values of 
$100-80,000$~rad and would imply a negligible degree of linear polarization from the flaring emission, which would be incompatible with the observations.

%For the observed wavelength of flaring emission (1.3~mm), even the lower limits on the electron density result in Faraday rotation values of $100-80,000$~rad and are not compatible with the observations.
This confirms that the X-ray-emitting flaring coronal loops (and their characteristics) are not compatible with the millimeter flares.

If we instead consider a scenario in which the linear polarization is produced in the upper corona and escapes to be observed, we can place limits on what electron densities are compatible with our observations.  The degree of  polarization  becomes vanishingly small when the product of the Faraday rotation per unit path length and the line of sight path length becomes larger than a value of several rad m$^{-1}$ \citep{cioffijones1980}.  If we assume that the large-scale magnetic field takes on a dipolar configuration, then for a base field strength of a few kG, the magnetic field strength takes on a value of 10~G near $r/R_{\star}\sim7$.  Making the same assumptions as above about the line of sight path length and $B_{||}$, the electron density needs to be $\lesssim10^{5}$~cm$^{-3}$ for AU Mic and $\lesssim5\times10^{6}$~cm$^{-3}$ for Proxima Cen in order to explain the observation of linear polarization as intrinsic to the stellar flares.  We note that no observational constraints exist for stellar coronae at these far distances above the stellar surface, but the underlying assumption of large loops is validated by recent application of dynamo models to stellar outer structures \citep{cohen2017}.

\subsection{Interpretation}
\label{sec:interpretation}

In order to interpret the origin of the six observed millimeter flares from M dwarfs, we assume that the magnetic geometry in these bursts is similar to what has been seen on the Sun and inferred for other stellar systems -- namely magnetic loops anchored to the star in starspots, and extending up into the tenuous stellar atmosphere.  The flares likely originate from a liberation of energy resulting from magnetic reconnection.  The short duration of the millimeter flares suggests that simple loop structures are involved.
Some fraction of this energy goes into particle acceleration and plasma heating, although we only have observational constraints on the former.   The unknowns include the geometry and length scales of the loops (i.e., one large loop-like structure or an arcade of loops), the magnetic field strength in the radio-emitting region, and the ambient electron density, although the linear polarization signal puts limits on $B_{||}$ and $n_{e}$.  Since the present observations are limited to millimeter wavelengths only, we do not know what the full spectral energy distribution of the radio emission is from microwave to millimeter wavelengths, and whether there are multiple contributing sources.  In this section, we consider incoherent emission from a population of accelerated particles, radiating either via gyrosynchrotron (\S\ref{sec:gyrosynchrotron}) or synchrotron (\S\ref{sec:synchrotron}) emission, and use the characteristics of the detected events and derived flare characteristics (described in \S\ref{sec:properties} and \S\ref{sec:flarechar} above) to establish likely scenarios.  We note that the conclusion in \S\ref{sec:trapping} that the energetic particles originate largely from directly precipitating particles (as opposed to mostly trapped particles) is also consistent with these low electron densities, as trapped particles with MeV energies require significantly higher electron densities to have the trapping time consistent with the timescale for collisional loss.

\subsubsection{Gyrosynchrotron}
\label{sec:gyrosynchrotron}

Gyrosynchrotron emission is the assumed mechanism for many solar flare emissions at microwave and millimeter frequencies \citep{white1992}.  This emission originates from mildly relativistic particles spiralling in the presence of magnetic fields.  The  gyrofrequency is $\nu_{B}$=2.8$\times$10$^{6}\times$B(G)~Hz with a range of harmonics, $s$, between 10 and 100.  For the observed frequency $\nu_{\rm obs} = s\nu_{B}$, this implies magnetic field strengths ranging from roughly $0.8-8$~kG. Electrons with energy $E=\gamma m_{e}c^{2}$ emit preferentially at frequencies $\nu\sim \gamma^{2} \nu_{B}$, for magnetic field strength, $B$, and Lorentz factor, $\gamma$.  For the above range of magnetic field strengths and observing frequency, this yields $\gamma$ from $18-180$.  The electron kinetic energy $E_{\rm kin}=(\gamma-1)m_{e}c^{2}$, with $m_{e}$ the mass of the electron and $c$ the speed of light, ranges from roughly $0.8-90$~MeV.  The negative spectral indices found for the two largest millimeter flares from AU Mic (A1) and Proxima Cen (P4) imply optically thin non-thermal emission.  Using the formulae for homogeneous sources \citep{Dulk:1985}, the derived particle distribution indices for these M dwarfs are consistent with a hard distribution of electrons ($\delta_\text{A1}=2.80\pm0.06$ and $\delta_\text{P4}=3.3\pm0.5$). Thus, there is a larger contribution from the most energetic electrons in these two events. 

Gyrosynchrotron emission in (non-solar) stellar radio flares has largely been studied at frequencies below about 10~GHz, far from the frequencies considered in this paper. Those investigations \citep[e.g.,][]{osten2005} have indicated a peak frequency varying from several GHz to $>$10~GHz. Table~\ref{tab:peak} lists the peak flux density of the spectral energy distribution for the A1 and P4 flares calculated using the observed spectral indices and fluxes for these two events.  We assume there is only one source of emission in this wavelength range and consider a range of possible peak frequencies between $10-200$~GHz.   For the smallest values of $\nu_{\rm peak}$, the predicted peak flux density for the P4 flare eclipses any recorded stellar flaring flux density measurement by orders of magnitude. We also record the integrated radio energy for this distribution for comparison, and note the significantly smaller range of E$_{R}$.

\begin{figure}[t]
    \centering
    \includegraphics[angle=90,scale=0.37]{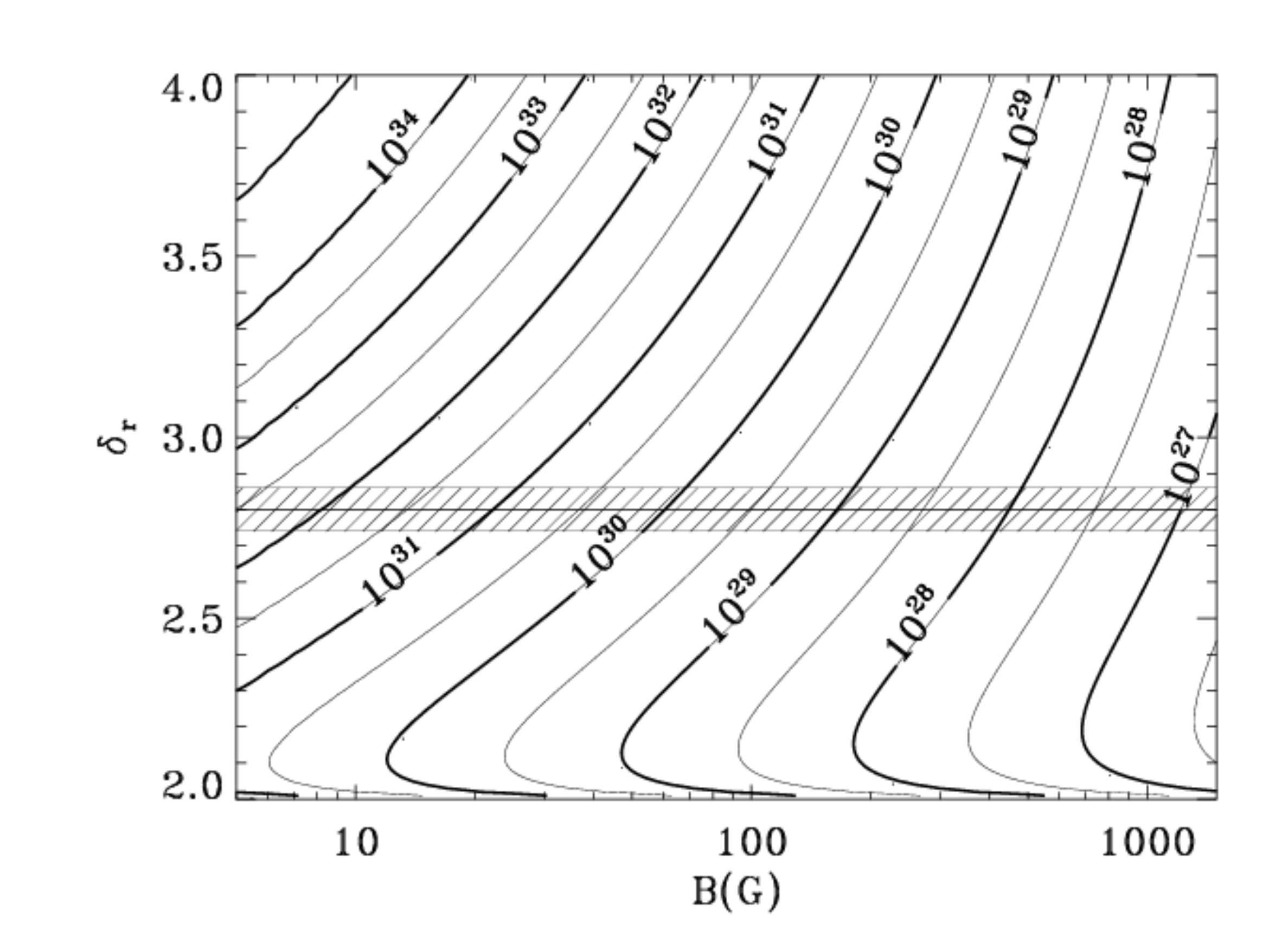}
     \includegraphics[angle=90,scale=0.37]{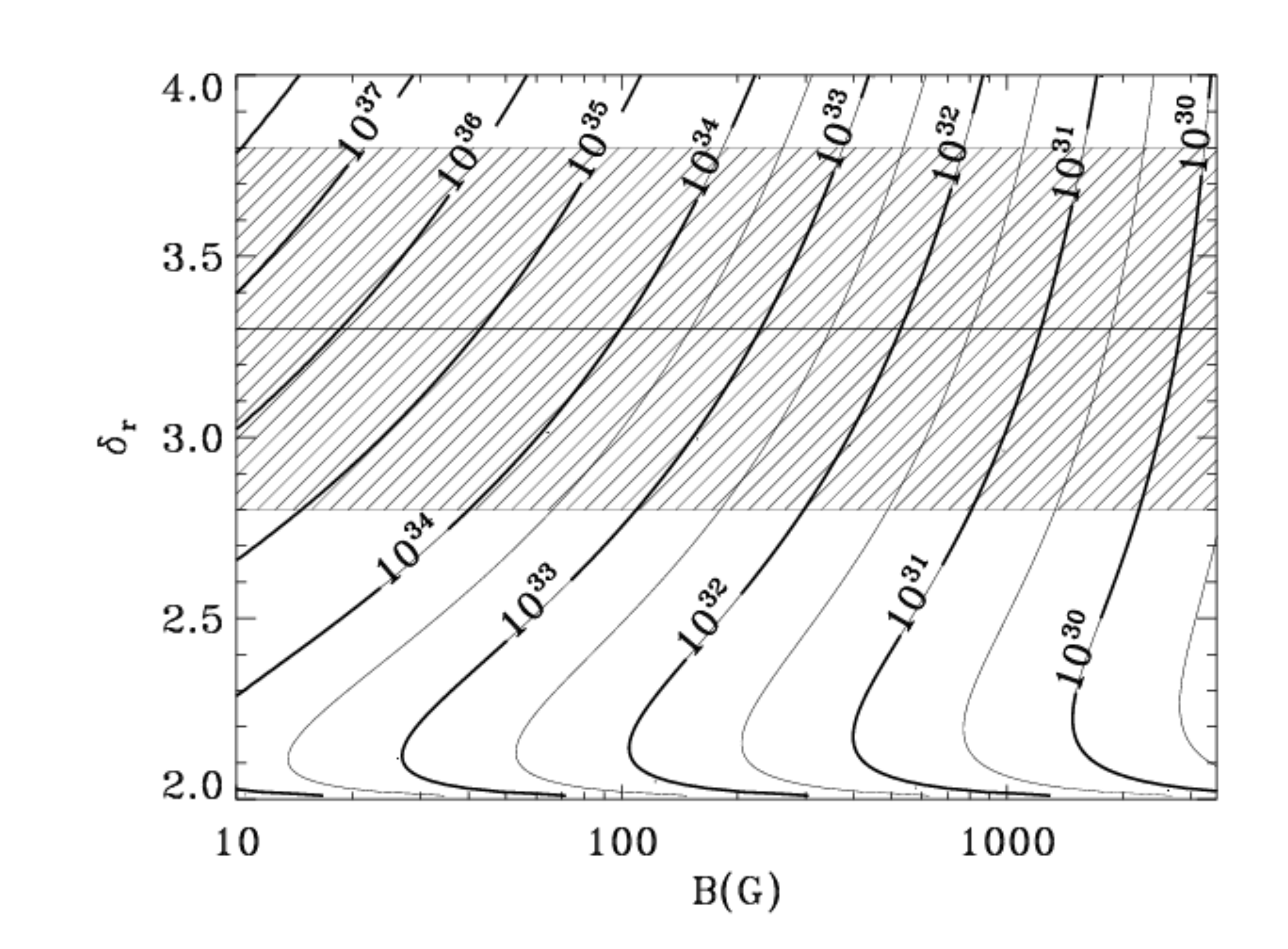}
    \caption{Contours of non-thermal energy, in units of erg, as a function of $\delta_{r}$, the index of the accelerated particle distribution, and the magnetic field strength in the radio-emitting source, for the A1 flare on AU Mic \emph{(left)} and the P4 flare on Proxima \emph{(right)}, under the assumption of gyrosynchrotron emission. Additional details are available in the text. The hatched lines indicate the range of $\delta_{r}$ constrained from the radio observations. }
    \label{fig:gyro_contours}
\end{figure}

If we assume that a single homogeneous emitting source is responsible for the microwave to millimeter spectral energy distribution, then we can make order of magnitude estimates for the non-thermal energy contained therein.  Figure~\ref{fig:gyro_contours} shows this dependence of non-thermal energy on the distribution of accelerated particles, $\delta_{r}$, and the magnetic field strength in the radio-emitting source, $B$, for the A1 flare from AU Mic (left) and the P4 flare from Proxima Cen (right), derived using the methodology established in \citet{smith2005} and \citet{Osten:2016} for gyrosynchrotron emissivities.  The peak frequency of the emission was assumed to be 50~GHz, although from inspecting Table~\ref{tab:peak} the energy can be within $1-2$ orders of magnitude for the different $\nu_{\rm pk}$ scenarios, and thus the contours are illustrative only.  A quiescent flux density of 390~$\mu$Jy for AU~Mic and 74~$\mu$Jy for Proxima Cen has been subtracted from the light curves before integrating.  The hatched lines indicate the range of $\delta_{r}$ derived from the observed spectral indices under the assumption of gyrosynchrotron emission.  The significantly smaller interval for A1 is a function of the smaller uncertainty on the spectral index, $\alpha$, derived using the more sensitive measurements from the full ALMA 12-m array (consisting of 37 12-m diameter antennas), rather than the Atacama Compact Array (ACA, consisting of 10 7-m diameter antennas).  If we  assert a causal connection between the non-thermal energy estimate and the minimum bolometric energy estimate from \S\ref{sec:bolometric}, this figure illustrates a range of magnetic field strengths in the radio-emitting plasma under the assumption of gyrosynchrotron emission.  For AU~Mic, the non-thermal energy is much less than the bolometric flare energy unless the magnetic field strength is less than a few hundred Gauss, while for Proxima Cen, the range of magnetic field strengths between a few hundred and a few thousand Gauss for gyrosynchrotron emission gives an approximate equipartition with the bolometric flare energy.

These field strengths are much higher than what was used in the calculations in Section~\ref{sec:magnetic} to explain the detection of linear polarization.  A value of $B$ roughly $100\times$ higher implies a value of the electron density lower by the same amount to
keep the same conditions on the escape of linearly polarized radiation.  This suggests densities of $\lesssim10^{3}$~cm$^{-3}$ for AU Mic and $\lesssim5\times10^{4}$~cm$^{-3}$ for Proxima Cen.  These densities are very small with respect to derived coronal densities
described above in \S\ref{sec:magnetic}.

\subsubsection{Synchrotron}
\label{sec:synchrotron}

Synchrotron emission is the favored explanation for both the linear polarization \citep{phillips1996} and 90~GHz flares observed on V773 Tau \citep{Massi:2006}.  Until now, there has not been another stellar radio flare with measured amounts of linear polarization.  As in the case of gyrosynchrotron emission, we expect that electrons will emit preferentially at frequencies where $\nu\sim \gamma^{2} B$.  Much higher harmonic numbers of the gyrofrequency are involved with synchrotron emission compared with gyrosynchrotron emission, thus favoring smaller magnetic field strengths.  The spectrum transitions from peaks at a small range of harmonic numbers to a more continuous broad emission spectrum.  Assuming a dipole field with a base magnetic field strength of a few kG, magnetic field strengths of 1, 10, and 100~G occur at $r/r_{0}\sim15$, $r/r_{0}\sim7$, and $r/r_{0}\sim3$, respectively. These values lead to $\gamma$ factors of 360, 113, and 36, respectively, or electron kinetic energies of 180, 57, and 18~MeV.   

As with gyrosynchrotron emission, the negative spectral indices observed for the millimeter flares presented here imply optically thin emission.  For a homogeneous, uniform, optically thin synchrotron source in \citet{Dulk:1985}, the particle distribution index is $\alpha = -(\delta_{r}-1)/2$.  The derived particle distribution indices for these M dwarf flares are slightly larger for synchrotron emission than for gyrosynchrotron emission,   $3.6\pm0.08$ (A1) and $4.5\pm0.7$ (P4) compared to 2.8 (A1) and 3.3 (P4), but are still indicative of a modestly hard electron distribution.  The expected linear polarization fraction is quite high, $r_{l}\sim0.8$ for both flares.  Depolarization of intrinsically linearly polarized emission could account for the difference between this expected value and the observed value. 

We have not computed an analog of Figure~\ref{fig:gyro_contours} using synchrotron emissivities, as there are too many unknowns.  \citet{Beasley:1998} noted an extreme flaring event on the active binary UX~Ari, which appeared to indicate gyrosynchrotron emission below a few tens of GHz and a spectrum rising  at 100~GHz  that could be indicative of a synchrotron component.  The small magnetic field strengths determined from our analyses suggest non-thermal energies far above the bolometric flare energies calculated from event rates, and imply orders of magnitude more energy going into accelerated particles.

\subsubsection{Synthesis and Implications}
\label{sec:synthesis}

It is likely that millimeter emission is a normal part of M dwarf flaring that we have missed until now.  The estimated event rate calculated from the six detected flares from AU Mic and Proxima Cen is entirely consistent with small to moderate energy flares expected to occur in the visible range as a normal consequence of stellar magnetic activity.  The two largest flaring events (A1 and P4 throughout the text) are characterized by short durations, falling spectral indices indicative of optically thin emission, and evidence for linear polarization.  Both of these events must involve a hard distribution of very energetic electrons, although the dominant emission mechanism (either gyrosynchrotron or synchroton) remains unclear from this small sample of events.  Given the linear polarization signal, synchrotron emission is favored, implying lower magnetic field strengths.  However, it is not possible to produce fast duration radio bursts along with a linear polarization signal via this mechanism from trapped electrons.  The former requires high electron densities, while the latter requires low electron densities-- a clear contradiction.  Thus, taken at face value, the characteristics of the observed events seem to rule out trapping of electrons.  

Millimeter emission has been detected previously from solar flares, often associated with extreme X-ray emission \citep{Krucker:2013}. These solar events have durations of a few minutes and positive spectral indices with $\alpha$ between $0.3-5$, opposite to what we measure during the M dwarf flares presented here.  The origin of solar millimeter emission is still debated, but synchrotron emission from ultra-relativistic electrons or an extreme-IR thermal extension of the heated chromosphere at 10,000~K are two favored possibilities.  It is clear, however, that all impulsive solar flares produce MeV electrons on prompt time scales, and the non-thermal population of energetic electrons producing millimeter emission is distinct from the population producing centimeter and hard X-ray emission.

Importantly, radio observations offer us the only opportunity to detect the signature of accelerated particles in flaring stellar atmospheres.  Unlike for the Sun, we cannot use hard X-ray observations due to a lack of sensitivity.  Even the largest hard X-ray stellar flares detected to date (with enhancements $1000\times$ the quiescent X-ray emission) are still consistent with thermal emission from a very hot thermal plasma at $100-300$~million~K \citep{Osten:2007,Osten:2010,Osten:2016,Pandey:2012}.  In this work, we have only considered the role of electrons in flaring emission.  However, protons and other energetic particles are also likely produced by these events and are of particular interest when considering the impact of stellar activity on planetary atmospheres.    

\pagebreak

\section{Conclusions}
\label{sec:conclusions}

It is clear that there is still much to be learned about M dwarfs, their activity, and the potential habitability of their planetary systems.  In this work, we have presented an analysis of six flares detected at millimeter wavelengths from AU Mic and Proxima Cen.  These are the first M dwarf flares detected in this wavelength regime.  The unique characteristics of these flares, especially the falling spectral indices and linear polarization observed at peak, suggest that the millimeter observations most likely trace optically thin synchrotron emission from precipitating electrons.  However, it is difficult to draw firm conclusions from such a small sample.  Stellar activity remains largely unexplored at millimeter wavelengths and offers potentially large payoffs in understanding particle acceleration in M dwarfs.

Going forward, multi-wavelength observing campaigns will be able to place important constraints on the accelerated particle population in stellar, especially M dwarf, flares.  ALMA will continue to be an important tool to explore the millimeter regime between $\sim80-900$~GHz with a particular strength at 1~mm or $\sim230$~GHz (ALMA Band 6).  Other millimeter facilities, such as the Submillimeter Array (SMA) and NOrthern Extended Millimeter Array (NOEMA), can also probe this wavelength regime and provide access to northern hemisphere targets.  To confirm the presence of linear polarization during M dwarf flaring events, future millimeter observations with full polarization calibration will be essential in order to determine whether any instrumental polarization has contributed to our previous results.  There is still a need to fill the observing gap between $\sim10-200$~GHz in order to better understand the processes occurring during stellar flares.  Sensitive next-generation radio facilities such as the ngVLA will partially address this, becoming excellent tools to study the $\sim10-100$~GHz spectral range.  Although these millimeter and radio observations alone will have immense value by solidifying the characteristics and rate of millimeter flares, the combination of these observations with monitoring at optical through X-ray wavelengths would provide additional constraints on flaring mechanisms and impacts on planetary habitability.  Indeed, all-sky surveys are already providing this kind of complementary information from the ground \cite[e.g., Evryscope,][]{Law:2016} and space \cite[e.g., TESS,][]{Ricker:2015}.

\vspace{1cm}
The authors gratefully acknowledge Samantha O'Sullivan and Cail Daley for their initial work with the AU Mic ALMA data that helped to spark the idea for the analysis presented here.  The authors would also like to thank Paolo Cortes for his helpfulness with ALMA's polarization calibration.  M.A.M. acknowledges support from a National Science Foundation Astronomy and Astrophysics Postdoctoral Fellowship under Award No. AST-1701406.  R.A.O. acknowledges support for this work from the Space Telescope Science Institute,  which is operated by the Association of Universities for Research in Astronomy, Inc., under NASA contract NAS 5-26555.  A.M.H. is supported by a Cottrell Scholar Award from the Research Corporation for Science Advancement.  This paper makes use of the following ALMA data: ADS/JAO.ALMA \#2012.1.00198.S and \#2016.A.00013.S. ALMA is a partnership of ESO (representing its member states), NSF (USA) and NINS (Japan), together with NRC (Canada) and NSC and ASIAA (Taiwan) and KASI (Republic of Korea), in cooperation with the Republic of Chile. The Joint ALMA Observatory is operated by ESO, AUI/NRAO and NAOJ. The National Radio Astronomy Observatory is a facility of the National Science Foundation operated under cooperative agreement by Associated Universities, Inc.

\bibliography{References.bib}

\pagebreak

\begin{deluxetable}{l c c c c c c}
\tablecolumns{5}
\tabcolsep0.1in\footnotesize
\tabletypesize{\small}
\tablewidth{0pt}
\tablecaption{Detected Millimeter Flare Properties \label{tab:prop}}
\tablehead{
\colhead{Star} &
\colhead{Flare$^{\dagger}$} & 
\colhead{Peak Flux Density} & 
\colhead{Peak $L_{R}$} &
\colhead{$t_{1/2}$} &
\colhead{$\alpha$} & 
\colhead{$|Q/I|$}
\\
\colhead{} & 
\colhead{} & 
\colhead{(mJy)} & 
\colhead{(10$^{13}$ erg s$^{-1}$ Hz$^{-1}$)} &
\colhead{(sec)} & \colhead{} & \colhead{} 
}
\startdata
AU Mic & A1 & 15  & 196 &35 &$-1.30\pm$0.05 & $>$0.12$\pm$0.04\\
 & A2 & 5 & 69 &9 & $^{\ddagger}$ & $^{\ddagger}$\\
\hline
Proxima Cen & P1 & 45 &  9.2&4 & $^{\ddagger}$ & $^{\ddagger}$\\
 & P2 & 20 & 4.1& 2.8  & $^{\ddagger}$ & $^{\ddagger}$\\
 & P3 & 10 &2.0 & 2.4 & $^{\ddagger}$ & $^{\ddagger}$ \\
 & P4 & 100 &20. & 16.4 &$-1.77\pm0.45$ & $>$0.19$\pm$0.02\\
\enddata
\tablecomments{$^{\dagger}$ Flares labeled A$n$ and P$n$ were detected from the flare stars AU~Mic and Proxima Cen, respectively. All flares are marked in Figure~\ref{fig:flarelc}.  $^{\ddagger}$ Not constrained by current observations.}
\end{deluxetable}

\begin{deluxetable}{c c c c c}
\tablecolumns{3}
\tabcolsep0.2in\footnotesize
\tabletypesize{\small}
\tablewidth{0pt}
\tablecaption{Flare Peak Flux Densities and Integrated Radio Energy\label{tab:peak}}
\tablehead{
\colhead{$\nu_{\rm peak}$} &
\colhead{SED Peak -- A1} & \colhead{$E_{r,int}$} &
\colhead{SED Peak -- P4} & \colhead{$E_{r,int}$} \\
\colhead{(GHz)} & 
\colhead{(mJy)} & \colhead{(erg)} &
\colhead{(mJy)} & \colhead{(erg)}
}
\startdata
10 & 840 & 10$^{24}$ & 26000 & 10$^{28}$\\
50 & 100 & 5$\times$10$^{23}$ & 1500 & 3$\times$10$^{27}$\\
100 & 42 &3$\times$10$^{23}$ & 450 & 2$\times$10$^{27}$\\
150 & 25 & 2$\times$10$^{23}$ & 220 & 8$\times$10$^{26}$\\
200 & 17 & 10$^{23}$ & 130 & 5$\times$10$^{26}$ \\
\enddata
\tablecomments{SED peak flux densities are calculated assuming gyrosynchrotron emission.  Flares are marked in Figure~\ref{fig:flarelc}.}
\end{deluxetable}

\end{document}